\documentclass[apj]{emulateapj}
\usepackage{mathptmx}

\def\gtorder{\mathrel{\raise.3ex\hbox{$>$}\mkern-14mu
             \lower0.6ex\hbox{$\sim$}}}
\def\ltorder{\mathrel{\raise.3ex\hbox{$<$}\mkern-14mu
             \lower0.6ex\hbox{$\sim$}}}

\slugcomment{Draft of \today}

\shorttitle{Type IIn SNe rise time}

\shortauthors{Ofek et al.}

\begin{document}




\title{Interaction-powered supernovae: Rise-time vs. peak-luminosity correlation and the shock-breakout velocity}
\author{Eran O. Ofek\altaffilmark{1},
Iair Arcavi\altaffilmark{1},
David Tal\altaffilmark{1},
Mark Sullivan\altaffilmark{2},
Avishay Gal-Yam\altaffilmark{1},
Shrinivas R. Kulkarni\altaffilmark{3},
Peter E. Nugent\altaffilmark{4,5},
Sagi Ben-Ami\altaffilmark{1},
David Bersier\altaffilmark{6},
Yi Cao\altaffilmark{3},
S. Bradley Cenko\altaffilmark{7},
Annalisa De Cia\altaffilmark{1},
Alexei V. Filippenko\altaffilmark{5},
Claes Fransson\altaffilmark{8},
Mansi M. Kasliwal\altaffilmark{9},
Russ Laher\altaffilmark{10},
Jason Surace\altaffilmark{10},
Robert Quimby\altaffilmark{11}, and
Ofer Yaron\altaffilmark{1},
}
\altaffiltext{1}{Benoziyo Center for Astrophysics, Weizmann Institute
  of Science, 76100 Rehovot, Israel}
\altaffiltext{2}{School of Physics and Astronomy, University of Southampton, Southampton SO17 1BJ, UK}
\altaffiltext{3}{Cahill Center for Astronomy and Astrophysics, California Institute of Technology, Pasadena, CA 91125, USA}
\altaffiltext{4}{Computational Cosmology Center, Lawrence Berkeley National Laboratory, 1 Cyclotron Road, Berkeley, CA 94720, USA}
\altaffiltext{5}{Department of Astronomy, University of California, Berkeley, CA 94720-3411, USA}
\altaffiltext{6}{Astrophysics Research Institute, Liverpool John Moores University, UK}
\altaffiltext{7}{Astrophysics Science Division, NASA Goddard Space Flight Center, Mail Code 661, Greenbelt, MD, 20771, USA}
\altaffiltext{8}{Department of Astronomy, The Oskar Klein Centre, Stockholm University, AlbaNova University Centre, SE-106 91 Stockholm, Sweden}
\altaffiltext{9}{Observatories of the Carnegie Institution for Science, 813 Santa Barbara St, Pasadena CA 91101 USA}
\altaffiltext{10}{Spitzer Science Center, MS 314-6, California Institute of Technology, Pasadena, CA 91125, USA}
\altaffiltext{11}{Kavli IPMU (WPI), the University of
Tokyo, 5-1-5 Kashiwanoha, Kashiwa-shi, Chiba, 277-8583, Japan}

\begin{abstract}

Interaction of supernova (SN) ejecta with the optically thick
circumstellar medium (CSM) of a progenitor star can result in a bright,
long-lived shock-breakout event.
Candidates for such SNe include Type IIn and
superluminous SNe.
If some of these SNe are powered by interaction,
then there should be a specific relation between their
peak luminosity, bolometric light-curve rise time,
and shock-breakout velocity.
Given that the shock velocity during
shock breakout is not measured, we expect a correlation,
with a significant spread, between the rise time and the peak luminosity
of these SNe.
Here, we present a sample of 15 SNe~IIn for which we have
good constraints on their rise time and peak luminosity
from observations obtained using the Palomar Transient Factory.
We report on a possible correlation between
the $R$-band rise time and peak luminosity of these SNe,
with a false-alarm probability of 3\%.
Assuming that these SNe are powered by interaction,
combining these observables and theory
allows us to deduce lower limits on the shock-breakout velocity.
The lower limits on the shock velocity we find are consistent
with what is expected for SNe (i.e., $\sim10^{4}$\,km\,s$^{-1}$).
This supports the suggestion that the early-time light curves
of SNe~IIn are caused by shock breakout in a dense CSM.
We note that such a correlation can arise
from other physical mechanisms.
Performing such a test on other classes of SNe
(e.g., superluminous SNe) can be used to rule out
the interaction model for a class of events.

\end{abstract}


\section{Introduction}
\label{sec:Introduction}

A supernova (SN) exploding within an optically thick circumstellar medium
(CSM) may have several unique characteristics.
First, if the Thomson optical depth in the CSM is larger than
$c/v_{{\rm s}}$, where $c$ is the speed of light and $v_{{\rm s}}$
is the shock velocity,
then the shock breakout will occur in the CSM rather than
near the stellar surface.
This will lead to shock-breakout events that are more luminous
and longer than those from normal supernovae
(SNe; e.g., Falk \& Arnett 1977; Ofek et al. 2010;
Chevalier \& Irwin 2011; Balberg \& Loeb 2011).

In a CSM with a slowly decreasing radial density profile
(e.g., a wind profile with density $\rho\propto r^{-2}$, where $r$ is the
radial distance),
the radiation-dominated shock will transform to a collisionless shock,
generating hard X-ray photons and TeV neutrinos
(Katz et al. 2011; Murase et al. 2011, 2013;
Ofek et al. 2013a).
While the collisionless shock traverses regions in which the 
Thomson optical depth, $\tau$, is above a few, the hard X-ray photons
can be converted to visible light (e.g., via comptonization;
Chevalier \& Irwin 2012; Svirski et al. 2012).
We refer to this as the optically thick interaction phase.
In most cases, emission of visible light from
the optically thick interaction phase will last on the order of 
ten times the shock-breakout time scale
(e.g., the time it takes the shock to
evolve\footnote{In a wind-profile CSM ($\rho_{{\rm CSM}}=K r^{-2}$) the optical depth is inversely
proportional to the radius.}
from $\tau \approx 30$ to $\tau \approx 3$).
Svirski et al. (2012) showed that the optically thick interaction
phase is characterized by bolometric emission with
a power-law or broken power-law light curve,
with specific power-law indices.
A recent example for such behavior was demonstrated
by Ofek et al. (2013d) for SN\,2010jl
(PTF\,10aaxf; see also Moriya et al. 2013; Fransson et al. 2013).
However, in most cases the shock-breakout time scale may be 
less than several days, and the optically thick interaction phase will thus be
short and hard to distinguish
in the optical band.
It is possible that later, when the interaction
is moving into the optically thin region, 
the hard X-ray photons traveling inward toward optically thick regions
(e.g., the cold dense shell; Chevalier \& Fransson 1994)
will be partially converted to optical photons.

Svirski et al. (2012) and Ofek et al. (2013d)
showed that for SNe having light curves
that are powered by interaction, there should exist
a specific relation between
the shock-breakout time scale,
the SN luminosity,
and the shock velocity at shock breakout.
For various reasons the shock velocity is hard to
measure.
Ignoring the shock velocity will introduce
considerable scatter into this relation.
However, we still expect a correlation,
with a significant spread,
between the SN rise time 
(i.e., a proxy for the shock-breakout time scale; Ofek et al. 2010)
and peak luminosity.

Type IIn SNe 
(e.g., Filippenko 1997)
are characterized by intermediate-width emission lines
which are presumably emitted by shock interaction
and/or recombination in optically thin gas in the CSM
due to the SN radiation field
(e.g., Chevalier \& Fransson 1994; Chugai 2001).
Furthermore, it was suggested that hydrogen-poor
superluminous SNe are powered by interaction
(Quimby et al. 2011; Chevalier \& Irwin 2011; see a review by Gal-Yam 2012),
as well as some other rare types of SNe
(e.g., Ben-Ami et al. 2013).

Here we perform a simple test of the interaction model
for SNe~IIn, by searching
for a correlation between the rise time and peak luminosity.
Indeed, we find a possible correlation between these properties.
However, we stress that
other models that can produce this correlation 
cannot yet be ruled out.
We present our SN sample and observations in \S\ref{sec:Observations},
and review the predictions in \S\ref{Pred}.
The data are analyzed in \S\ref{Analysis}, and
we discuss the results in \S\ref{Disc}.

\section{Sample and observations}
\label{sec:Observations}

The Palomar Transient Factory
(PTF\footnote{http://www.astro.caltech.edu/ptf/};
Law et al. 2009; Rau et al. 2009)
and its extension the intermediate PTF (iPTF)
found over 2100 spectroscopically confirmed SNe.
We selected 19 SNe~IIn for which
PTF/iPTF has good coverage of the light-curve rise and peak;
they are listed in Table~\ref{tab:sample}.
Optical spectra were obtained with a variety of telescopes and
instruments, including the Double Spectrograph (Oke \& Gun 1982) at the 
Palomar 5-m Hale telescope, the Kast spectrograph (Miller \& Stone 1993)
at the Lick 3-m Shane telescope, the Low Resolution Imaging Spectrometer
(Oke et al. 1995) on the Keck-1 10-m telescope, and the
Deep Extragalactic Imaging Multi-Object Spectrograph (Faber et al. 2003)
on the Keck-2 10-m telescope. 
A representative spectrum of each SN is available through the WISeREP
website\footnote{http://www.weizmann.ac.il/astrophysics/wiserep/}
(Yaron \& Gal-Yam 2012).
\begin{figure*}
\centerline{\includegraphics[width=17cm]{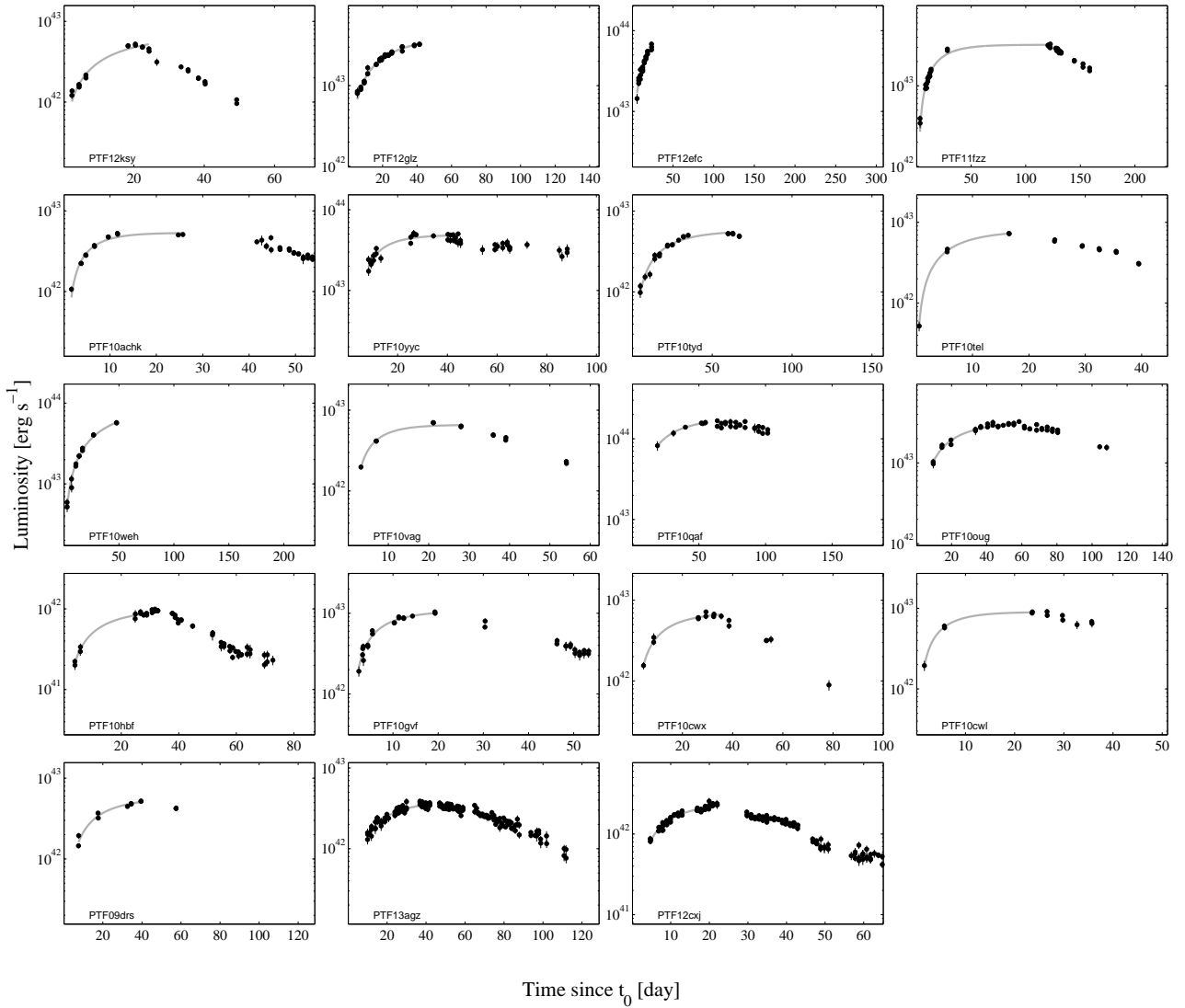}}
\caption{The light curves of the 19 SNe in our sample,
for which we attempted to fit the rise time using
the exponential rise function (Eq.~\ref{eq:rise_exp}).
The black circles show the PTF/iPTF $R$-band measurements with
their uncertainties, while the gray lines represent the best-fit
exponential rise function.
The SN name is marked in each subplot.
The values of $t_{0}$ are listed in Table~\ref{tab:sample}.
\label{fig:AllLC}}
\end{figure*}
%
%
\begin{deluxetable*}{lrrllllllllccc}
\tablecolumns{14}
\tablewidth{0pt}
\tablecaption{Supernovae Sample}
\tablehead{
\colhead{Name}              &
\colhead{RA}                &
\colhead{Dec}               &
\colhead{$z$}                 &
\colhead{DM}                &
\colhead{$E_{B-V}$}    &
\colhead{$t_{0}$}            &
\colhead{$t_{{\rm max}}$}  &
\colhead{$L_{{\rm max}}$}  &
\colhead{$L_{0}$}           &
\colhead{$t_{{\rm e}}$}     &
\colhead{$\chi^{2}$/dof}   &
\colhead{$v_{{\rm bo}}$}    &
\colhead{$\log_{10}{K}$}             \\
\colhead{}                  &
\colhead{deg}               &
\colhead{deg}               &
\colhead{}                  &
\colhead{mag}               &
\colhead{mag}               &
\colhead{MJD}               &
\colhead{MJD}               &
\colhead{erg\,s$^{-1}$}     &
\colhead{erg\,s$^{-1}$}     &
\colhead{day}               &
\colhead{}                 &
\colhead{km\,s$^{-1}$}      &
\colhead{[g\,cm$^{-1}$]}
}
\startdata
       PTF10cwx &   188.3189 &$    -0.0530$ & 0.073 & 37.58 & 0.025 & 55243.3 &   55275.3 &$ 7.2\times10^{42}$ &$ 4.7\times10^{44}$ &$  12.7 \pm   5.7$ &  8.5/7  &  3300 &  17.0\\ 
       PTF10gvf &   168.4385 &$    53.6291$ & 0.081 & 37.82 & 0.011 & 55319.9 &   55337.5 &$ 1.0\times10^{43}$ &$ 5.5\times10^{44}$ &$   6.6 \pm   1.5$ & 27.4/15 &  4600 &  16.7\\ 
       PTF10hbf &   193.1944 &$    -6.9220$ & 0.042 & 36.32 & 0.036 & 55292.1 &   55320.6 &$ 9.9\times10^{41}$ &$ 6.5\times10^{43}$ &$  13.3 \pm   4.4$ &  3.9/9  &  1700 &  17.1\\ 
       PTF10oug &   260.1866 &$    29.0738$ & 0.150 & 39.26 & 0.042 & 55378.5 &   55424.4 &$ 3.3\times10^{43}$ &$ 2.5\times10^{45}$ &$  20.9 \pm   5.3$ & 13.8/14 &  4600 &  17.2\\ 
       PTF10qaf &   353.9287 &$    10.7758$ & 0.284 & 40.82 & 0.074 & 55350.4 &   55409.4 &$ 1.7\times10^{44}$ &$ 1.3\times10^{46}$ &$  25.6 \pm   8.7$ &  0.3/4  &  7400 &  17.3\\ 
       PTF10tyd &   257.3309 &$    27.8191$ & 0.063 & 37.25 & 0.065 & 55419.5 &   55470.5 &$ 5.3\times10^{42}$ &$ 4.0\times10^{44}$ &$  20.4 \pm   2.0$ & 34.9/17 &  2500 &  17.2\\ 
       PTF10vag &   326.8270 &$    18.1310$ & 0.052 & 36.81 & 0.111 & 55445.4 &   55464.9 &$ 7.0\times10^{42}$ &$ 3.6\times10^{44}$ &$   5.6 \pm   2.5$ & 16.4/4  &  4300 &  16.7\\ 
       PTF10weh &   261.7103 &$    58.8521$ & 0.138 & 39.06 & 0.032 & 55450.3 &   55526.3 &$ 5.8\times10^{43}$ &$ 5.8\times10^{45}$ &$  54.8 \pm  12.4$ & 14.8/11 &  4000 &  17.7\\ 
       PTF10yyc &    69.8221 &$    -0.3488$ & 0.214 & 40.12 & 0.046 & 55476.7 &   55506.8 &$ 5.1\times10^{43}$ &$ 3.0\times10^{45}$ &$   9.1 \pm   3.7$ & 26.5/14 &  7100 &  16.9\\ 
      PTF10achk &    46.4898 &$   -10.5225$ & 0.033 & 35.77 & 0.063 & 55534.3 &   55551.5 &$ 5.3\times10^{42}$ &$ 2.6\times10^{44}$ &$   5.0 \pm   1.0$ & 54.8/10 &  4000 &  16.6\\ 
       PTF11fzz &   167.6945 &$    54.1052$ & 0.082 & 37.85 & 0.011 & 55723.6 &   55798.6 &$ 3.3\times10^{43}$ &$ 2.4\times10^{45}$ &$  18.2 \pm   1.4$ & 121.1/19& 4800 &  17.2\\ 
       PTF12cxj &   198.1612 &$    46.4851$ & 0.036 & 35.96 & 0.011 & 56029.6 &   56050.1 &$ 2.6\times10^{42}$ &$ 1.5\times10^{44}$ &$   9.1 \pm   1.0$ &  93.2/51& 2600 &  16.9\\ 
       PTF12glz &   238.7210 &$     3.5354$ & 0.079 & 37.76 & 0.131 & 56107.6 &   56155.4 &$ 3.3\times10^{43}$ &$ 2.7\times10^{45}$ &$  26.3 \pm   2.2$ &  42.7/33& 4300 &  17.3\\ 
       PTF12ksy &    62.9421 &$   -12.4669$ & 0.031 & 35.66 & 0.043 & 56232.5 &   56256.7 &$ 5.2\times10^{42}$ &$ 3.7\times10^{44}$ &$  17.3 \pm   6.4$ & 136.2/12& 2700 &  17.2\\ 
      iPTF13agz &   218.6338 &$    25.1621$ & 0.057 & 37.02 & 0.033 & 56377.5 &   56418.3 &$ 3.9\times10^{42}$ &$ 2.8\times10^{44}$ &$  18.0 \pm   2.5$ &  76.7/58& 2400 &  17.2\\ 
\hline
       PTF09drs &   226.6257 &$    60.5943$ & 0.045 & 36.49 & 0.017 & 55025.5 &   55066.5 &$ 5.2\times10^{42}$ &$ 3.7\times10^{44}$ &$  17.7 \pm  16.3$ &  14.3/6 & 2600 &  17.2\\ 
       PTF10cwl &   189.0919 &$     7.7939$ & 0.085 & 37.93 & 0.022 & 55245.1 &   55261.7 &$ 9.1\times10^{42}$ &$ 4.5\times10^{44}$ &$   5.0 \pm  12.8$ &   0.6/2 & 4800 &  16.6\\ 
       PTF10tel &   260.3778 &$    48.1298$ & 0.035 & 35.93 & 0.016 & 55427.8 &   55442.6 &$ 7.3\times10^{42}$ &$ 3.9\times10^{44}$ &$   6.6 \pm   3.5$ &   4.3/2 & 4100 &  16.7\\ 
       PTF12efc &   224.1447 &$    39.6848$ & 0.234 & 40.34 & 0.012 & 56052.7 &   56155.5 &$ 6.8\times10^{43}$ &$ 7.9\times10^{45}$ &$  88.0 \pm$\nodata&  18.9/19& 3700 &  17.9 
\enddata
\tablecomments{The sample of SNe~IIn. DM (mag) is the distance modulus of the SN host galaxy.
$E_{B-V}$ is the Galactic extinction in the SN direction (Schlegel et al. 1998),
$t_{0}$ is the MJD of the fitted zero flux,
$t_{{\rm max}}$ is the MJD of the $R$-band light-curve peak,
and $L_{{\rm max}}$ is the corresponding peak luminosity.
$L_{0}= L_{{\rm max}}(t/t_{{\rm bo}})^{-\alpha}$, where the time is measured in seconds (e.g., Eq.~\ref{eq:L}).
$t_{{\rm e}}$ is the exponential rise time of the early-time light curve, and $v_{{\rm bo}}$ is the
lower limit on the shock velocity deduced from Eq.~\ref{eq:vbo2} and assuming $\epsilon=0.3$, $w=2$, and $m=10$;
$\kappa=0.34$\,cm$^{2}$\,g$^{-1}$.
The mass-loading parameter $K=\dot{M}/(4\pi v_{{\rm w}})$ is calculated assuming a wind profile with $w=2$.
We assumed that the relative error in $L_{{\rm max}}$ is 20\%.
SNe below the horizontal line have relative errors in $t_{{\rm e}}$ larger
than 50\% and were excluded from our correlation analysis.
{\bf Information for Individual objects:}\\
PTF\,09drs -- Ofek et al. (2013a).\\
PTF\,10cwl (CSS100320:123622+074737) -- Drake et al. (2010).\\
PTF\,10tel (SN\,2010mc) -- Ofek (2012); Ofek et al. (2013a); Ofek et al. (2013b).\\
PTF\,10weh -- Ben-Ami et al. (2010); Ofek et al. (2014b).\\
PTF\,12efc -- is a candidate Type Ia SN interacting with its CSM (Silverman et al. 2013).\\
PTF\,12cxj -- Ofek et al. (2014b).\\
PTF\,12glz -- Gal-Yam et al. (2012).\\
PTF\,10cwx, PTF\,10gvf, PTF\,10hbf, PTF\,10oug, PTF\,10qaf, PTF\,10tyd,
PTF\,10vag, PTF\,10yyc, PTF\,10achk, PTF\,11fzz, PTF\,12ksy, iPTF\,13agz --
Reported here for the first time.\\
PTF\,10cwx -- A spectrum of this SN obtained during
maximum light shows narrow Balmer emission lines,
and possible weak He\,I lines, with a moderately blue continuum.
Another spectrum taken about seven weeks after maximum light
still shows a blue continuum with narrow Balmer emission lines.\\
PTF\,10gvf -- The first spectrum, taken about two weeks
prior to maximum light, shows Balmer as well as He\,I and He\,II
emission lines. A week later, the spectrum becomes
bluer, but the He lines are not detected. Seven weeks after maximum light,
the Balmer lines are still strong and become wider.\\
PTF\,10hbf -- A spectrum taken about two weeks after maximum light
shows an intermediate-width H$\alpha$ line.\\
PTF\,10oug -- A single spectrum of this SN taken about 27\,days
prior to maximum light shows a blue continuum with Balmer emission lines.\\
PTF\,10qaf -- A series of spectra taken from maximum light until
about three months after maximum light show Balmer emission lines.
The H$\beta$ line develops a weak P-Cygni profile about one month
after maximum light.\\
PTF\,10tyd -- The first spectrum was obtained about 26\,days
prior to peak luminosity. It exhibits a blue continuum
with intermediate-width Balmer lines.
A spectrum taken about one month after maximum light is very
similar to the first spectrum.\\
PTF\,10vag -- The first spectrum, obtained about 10\,days prior
to peak luminosity, shows a blue continuum with Balmer and He\,I
emission lines. The He\,I lines are still visible about 20\,days
after maximum light.\\
PTF\,10yyc -- A spectrum taken during maximum light shows a
blue continuum with Balmer emission lines.\\
PTF\,10achk -- Two spectra taken at maximum light and 10\,days
later shows a blue continuum with Balmer emission lines.
The first spectrum also shows He\,I emission lines.\\
PTF\,11fzz -- The first spectrum was taken during the SN rise,
about 10\,days after discovery. This spectrum shows a
blue continuum with Balmer and He\,I intermediate-width
emission lines. 75\,days after maximum light, the
strong Balmer emission lines are still present.\\
PTF\,12ksy -- The first spectrum, obtained about 20\,days
prior to peak luminosity, shows a blue continuum
with Balmer and He\,I emission lines. One month later,
the H$\alpha$ emission line exhibits a narrow absorption at
velocity of about $-500$\,km\,s$^{-1}$,
while the He\,I $\lambda$5876 line develops a strong P-Cygni profile.\\
iPTF\,13agz -- A spectrum obtained about one month after maximum light
shows a blue continuum with Balmer emission lines.
}
\label{tab:sample}
\end{deluxetable*}

The PTF/iPTF data were reduced using the IPAC
pipeline (Laher et al. 2014).
The photometric calibration is described by Ofek et al. (2012a).
The photometry was performed by running point-spread-function (PSF) fitting
on subtracted images
(e.g., Ofek et al. 2013c), and
the photometric measurements of all SNe in our sample
are listed in Table~\ref{tab:phot} and shown in  Figure~\ref{fig:AllLC}.
%
\begin{deluxetable}{llllll}
\tablecolumns{6}
\tablewidth{0pt}
\tablecaption{Supernovae photometry}
\tablehead{
\colhead{Name}              &
\colhead{Telescope}         &
\colhead{Filter}            &
\colhead{MJD}               &
\colhead{$R_{{\rm PTF}}$}     &
\colhead{Err}               \\
\colhead{}                  &
\colhead{}                  &
\colhead{}                  &
\colhead{day}               &
\colhead{mag}               &
\colhead{mag}               
}
\startdata
  PTF12ksy & PTF       & R   &  56202.486  & 20.414 &  0.143 \\ 
  PTF12ksy & PTF       & R   &  56235.251  & 19.175 &  0.098 \\ 
  PTF12ksy & PTF       & R   &  56235.280  & 19.038 &  0.059 \\ 
  PTF12ksy & PTF       & R   &  56237.255  & 18.840 &  0.057 \\ 
  PTF12ksy & PTF       & R   &  56237.288  & 18.919 &  0.033 
\enddata
\tablecomments{Photometric measurements of the SNe in the sample.
This table contains measurements from the PTF/iPTF telescope
as well as the Palomar 60-inch and Liverpool 2-m telescope.
It is published in its entirety in the electronic edition of
{\it ApJ}. A portion of the full table is shown here for
guidance regarding its form and content.}
\label{tab:phot}
\end{deluxetable}

\section{Predictions}
\label{Pred}

Next we briefly review the
predictions regarding the peak luminosity
and rise time in the context of the interaction model.
%
Ofek et al. (2014a) used the Chevalier (1982)
self-similar hydrodynamical solution
describing ejecta with a power-law velocity distribution
propagating through a CSM with a power-law density distribution.
Combining this with the shock-breakout properties
and assuming conversion of kinetic energy into luminosity,
Ofek et al. predicted a relation of the form
\begin{equation}
v_{{\rm bo}} = t_{{\rm bo}}^{(\alpha-1)/3}  \Big[ 2\pi \epsilon \frac{m-w}{m-3} (w-1)\frac{c}{\kappa L_{0}} \Big]^{-1/3}.
\label{eq:vbo2}
\end{equation}
Here $v_{{\rm bo}}$ is the shock-breakout velocity,
$t_{{\rm bo}}$ is the shock-breakout time scale,
$\epsilon$ is the efficiency of converting the kinetic
energy to luminosity,
$\kappa$ is the CSM opacity,
$m$ is the power-law index of the ejecta velocity
distribution,
and $w$ is the negative power-law index of the radial density
profile of the CSM (i.e., $\rho = Kr^{-w}$).
We note that for a wind profile CSM ($w=2$), 
the mass-loading parameter is $K=\dot{M}/(4\pi v_{{\rm w}})$, where
$\dot{M}$ is the mass-loss rate and $v_{{\rm w}}$ is the CSM velocity.
$L_{0}$ is the luminosity extrapolated to a
time of 1\,s and is defined by the relations
\begin{equation}
L = L_{0} t^{\alpha},
\label{eq:L}
\end{equation}
and
\begin{equation}
\alpha=\frac{(2-w)(m-3)+3(w-3)}{m-w}.
\label{eq:alpha}
\end{equation}

To summarize, in interaction-powered SNe
we expect a relation between
$t_{{\rm bo}}$, $v_{{\rm bo}}$, and $L_{0}$ (Eq.~\ref{eq:vbo2}).
The relevant observables are the rise time,
which is a proxy for the shock-breakout time scale
(e.g., Ofek et al. 2010),
and the peak luminosity, which is a function
of $L_{0}$ and $t_{{\rm bo}}$.
Therefore, we expect a correlation between the SN peak luminosity
and its rise time.
However,
given the relatively large power-law index in which $v_{{\rm bo}}$
appears in Equation~\ref{eq:vbo2},
relative to those of $t_{{\rm bo}}$ and $L_{0}$
($v_{{\rm bo}}^{3}\propto t_{{\rm bo}}^{\alpha-1}L_{0}$),
we predict that this correlation will have a large spread
(i.e., the correlation will be weak rather than tight).

\section{Analysis}
\label{Analysis}

In the context of the CSM-shock-breakout model,
characterization of the SN rise time requires a model
for the functional shape of the rising light curve.
Although some progress has been made (e.g., Ginzburg \& Balberg 2014),
we still lack such a model.
For simplicity, here we fit each SN rising light curve
with an exponential function of the form
\begin{equation}
L = L_{{\rm max}}\{1-\exp{[(t_{0} - t)/t_{{\rm e}}]}\}.
\label{eq:rise_exp}
\end{equation}
Here $L$ is the luminosity at time $t$,
and the free parameters in the fit are
the peak luminosity $L_{{\rm max}}$,
the time when the flux is zero $t_{0}$,
and the characteristic rise time $t_{{\rm e}}$.
We note that, in our analysis, instead of using the fitted $L_{{\rm max}}$,
we used the actual maximum observed luminosity.
This was done in order to avoid the effect of a possible
covariance between $t_{{\rm e}}$ and $L_{{\rm max}}$ that
may arise from the fitting process.
The $R$-band luminosity was calculated
taking into account the Galactic extinction
in the SN direction
(Cardelli et al. 1989; Schlegel et al. 1998), the SN luminosity distance
(WMAP3 cosmology; Spergel et al. 2007),
and assuming that the absolute magnitude of the Sun in the $R_{{\rm PTF}}$ band
is 4.66\,mag (Ofek et al. 2012a).

We also attempted to fit a $t^{2}$ law of the form
\begin{equation}
L = L_{{\rm max}} \Big( 1 - \Big[ \frac{t-t_{{\rm max}}}{\Delta{t}} \Big]^{2} \Big).
\label{eq:rise_t2}
\end{equation}
Here $t_{{\rm max}}$ is the maximum of the parabolic fit,
and $\Delta{t}$ is the time from zero to maximum luminosity.
In this case, the characteristic rise time
(i.e., the time it takes the light curve to rise by a factor of $\exp{[1]}$)
is given by
\begin{equation}
t_{{\rm e}} =  \Delta{t}(1 + \sqrt{1-e^{-1}}).
\label{eq:te_t2}
\end{equation}

\begin{figure}
\centerline{\includegraphics[width=8.5cm]{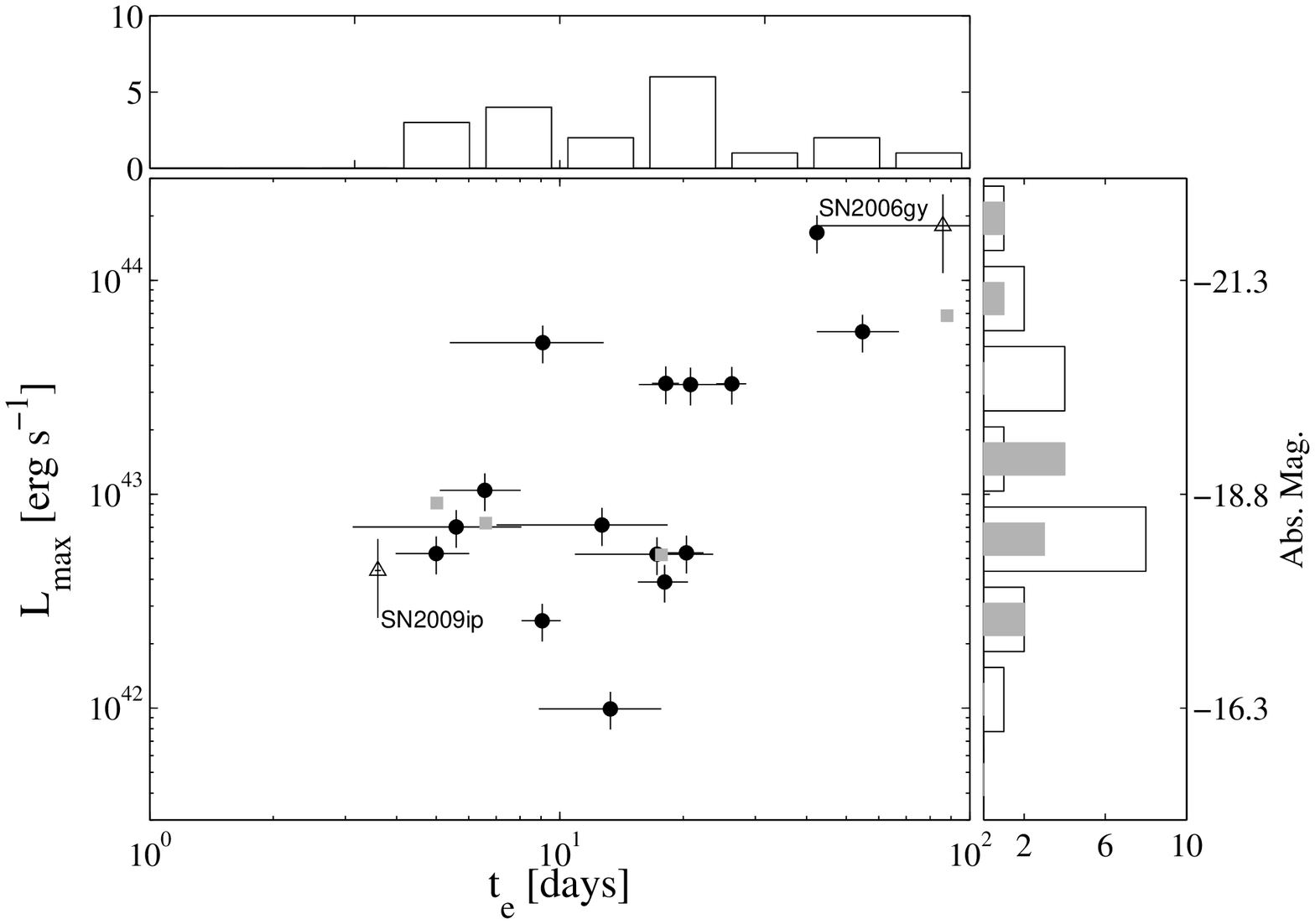}}
\caption{Peak luminosity ($L_{{\rm max}}$) vs. exponential rise ($t_{{\rm e}}$)
for the SNe in our sample.
The black circles are for the SNe whose
relative errors on the best-fit
exponential rise are smaller than $50\%$,
while the gray squares are for all the other SNe.
We did~not plot the errors for the gray squares,
and the corresponding SNe were not used in our correlation analysis.
The white-face histograms on the upper and right sides
present the $t_{{\rm e}}$ and $L_{{\rm max}}$ distributions,
respectively, for all 19 sources.
The narrower gray bars on the right-hand histogram
show the peak absolute magnitude distribution
of 11 SNe~IIn discussed by Kiewe et al. (2012).
Also shown (empty triangles) are the positions of some
other events: SN\,2006gy (Ofek et al. 2007; Smith et al. 2007)
and SN\,2009ip (Prieto et al. 2013; Ofek et al. 2013c; Margutti et al. 2014).
The rise time for these SNe was fitted in a way similar to
that for the main SNe in our sample.
\label{fig:te_Lp_corr}}
\end{figure}

Both fits provide a reasonable empirical
description of the rising light curves
(Fig.~\ref{fig:te_Lp_corr}).
For the purpose of the analysis presented here, we use the
$t_{{\rm e}}$ obtained from the exponential fit (Eq.~\ref{eq:rise_exp}).
We note that, qualitatively,
the results do~not change if one
uses the rise time obtained from the parabolic fit.
The best-fit exponential rise time and maximum luminosity
for each SN are listed in Table~\ref{tab:sample}.

For the 19 SNe in our sample,
we estimated the errors in the rise times using the
bootstrap technique (Efron 1982).
For 4 out of 19 events the
relative errors are larger than 50\%; they
appear in Table~\ref{tab:sample} below the horizontal line
and are marked on the plots with gray boxes.
We flagged these events as unreliable and they were not used
in the correlation analysis.

Figure~\ref{fig:te_Lp_corr} presents the 
observed $L_{{\rm max}}$ vs. $t_{{\rm e}}$.
The Spearman rank correlation of the remaining 15 SNe in our sample
is 0.49. Using the bootstrap technique (Efron 1982)
we find that the probability to get a correlation coefficient
larger than that is 0.03.
Therefore, the correlation is significant at the 2.5-$\sigma$ level.
We note that the Spearman rank correlation is not sensitive
to the distribution of variables, while the use of the bootstrap
technique give us an estimate of the false-alarm probability
taking into account the scatter in the data, but
without using the formal errors in the variables.
Furthermore, we tried different statistical approaches
that make quite different assumptions and obtained very similar results
(e.g., Brandon 2007).

Our results may be affected by selection biases
and therefore should be treated with care.
A possible selection effect is that SNe with longer rise times
are easier to detect even if they are faint.
However, such a selection effect will introduce
an anticorrelation between the rise time and peak luminosity.
Another concern is if the luminosity span of
the light-curve rise can affect our fitting.
In order to check for this and other selection effects, we also
look for correlations between the luminosity ratio
of the first SN detection and its peak luminosity,
and the SN rise time as well as the SN peak luminosity.
We do not find any evidence for such correlations.

We conclude that there is marginal evidence for a correlation
between the rise time and peak luminosity of SNe~IIn,
and that in this stage we cannot rule out
the possibility that these SNe are powered by interaction.

\begin{figure}
\centerline{\includegraphics[width=8.5cm]{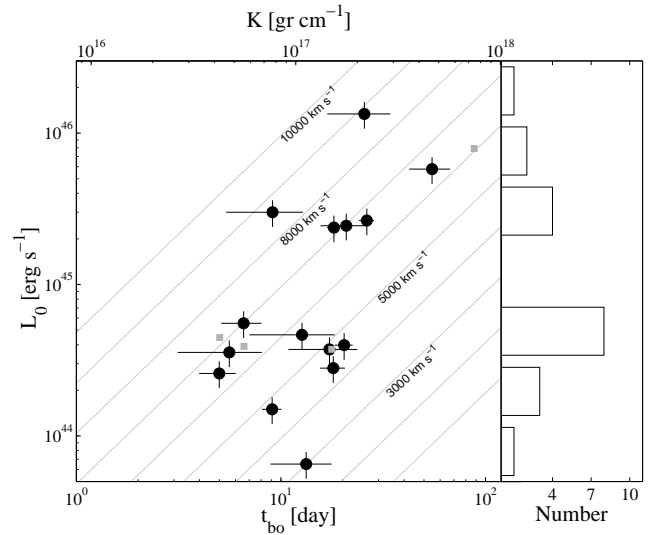}}
\caption{Shock-breakout time scale ($t_{{\rm bo}}$)
vs. $L_{0}$.
The shock breakout time scale is assumed to
be identical to $t_{{\rm e}}$,
while $L_{0}$ is calculated from Equation~\ref{eq:L}.
Symbols as in Figure~\ref{fig:te_Lp_corr}.
Lines of equal shock-breakout velocity,
as calculated using Equation~\ref{eq:vbo2}, are shown
as gray dashed lines.
We note that, in the context of the interaction model,
our measurements provide a lower limit
on $L_{0}$ and $t_{{\rm bo}}$; thus, they
they represent a lower limit on the breakout shock velocity.
The upper abscissa gives the mass-loading parameter $K$
assuming a wind profile (i.e., $w=2$; see Eq. 8 in Ofek et al. 2014a).
The vertical histogram on the right shows the $L_{0}$ distribution
for all 19 SNe in our sample.
\label{fig:IIn_PeakL_RiseTime}}
\end{figure}

Next, we use Equation~\ref{eq:L}, with the
constants listed below, to calculate $L_{0}$.
Figure~\ref{fig:IIn_PeakL_RiseTime} shows
$L_{0}$ vs. $t_{{\rm e}}$.
$L_{0}$ is a function of $L_{{\rm max}}$ and $t_{{\rm e}}$,
and therefore Figure~\ref{fig:IIn_PeakL_RiseTime} shows
two nonindependent parameters.
However, in Figure~\ref{fig:te_Lp_corr} we
already showed that
there is a possible correlation between an independent
version of these parameters.
We note that in the context of the interaction model
$t_{{\rm e}}$ is our best estimate for $t_{{\rm bo}}$.
Most importantly,
the power-law index of $v_{{\rm bo}}$ in Equation~\ref{eq:vbo2}
is larger than that of $L_{0}$ and $t_{{\rm bo}}$.
Therefore, we expect that Figure~\ref{fig:IIn_PeakL_RiseTime}
will exhibit a large scatter.

We stress that theory as well as
some UV observations suggest that the
bolometric rise time can be faster than the $R$-band
rise time (e.g., Roming et al. 2012;
Gal-Yam et al. 2013);
hence, our $t_{{\rm e}}$ is likely only an upper limit on
$t_{{\rm bo}}$.
Moreover, $L_{0}$ was estimated based on the $R$-band magnitude
rather than the bolometric magnitudes.
Therefore, these $L_{0}$ values should be regarded as lower limits.

Overplotted on Figure~\ref{fig:IIn_PeakL_RiseTime}
are the equal shock-velocity contours, as calculated
using Equation~\ref{eq:vbo2},
assuming $w=2$ (i.e., wind profile), $m=12$,
$\kappa=0.34$\,cm$^{2}$\,r$^{-1}$,
and $\epsilon=0.3$.
These values of $m$ and $w$ were also used to calculate $L_{0}$.
We note that $m=12$ (10) is expected in the case of a convective
(radiative) envelope (Matzner \& McKee 1999),
and that the value of $\alpha$ is not very sensitive to the value of $m$
(see Ofek et al. 2014a).
Furthermore, $\epsilon=0.3$ is close to the maximum possible
efficiency.
Given that our measurements provide an upper limit on $t_{{\rm bo}}$
and a lower limit on $L_{0}$,
the deduced breakout shock velocities in Figure~\ref{fig:IIn_PeakL_RiseTime}
are only a lower limit on the actual shock velocity at breakout.

\section{Discussion}
\label{Disc}

There is a growing line of evidence that SNe~IIn
are embedded in a large amount of CSM ejected
months to years prior to their explosions
(e.g., Dopita et al. 1984;
Weiler et al. 1991;
Chugai \& Danziger 1994;
Smith et al. 2008;
Gal-Yam \& Leonard 2009;
Kiewe et al. 2012;
Ofek et al. 2013c).
In some cases we probably see optical outbursts
associated with these mass-loss events
(e.g., Foley et al. 2007; Pastorello et al. 2007;
Mauerhan et al. 2012;
Corsi et al. 2013;
Fraser et al. 2013;
Ofek et al. 2013b; 2014b).
This CSM is likely to be optically thick
and lead to luminous and long shock-breakout events
(Ofek et al. 2010; Chevalier \& Irwin 2011).

For some SNe the early-time light curve is powered by shock breakout
in a dense CSM followed by conversion of the kinetic
energy to optical luminosity via shock interaction
in optically thick regions.
In such cases,
Svirski et al. (2012) and Ofek et al. (2014a) predicted
a relation between the shock-breakout time
scale ($t_{{\rm bo}}$), velocity ($v_{{\rm bo}}$),
and the SN peak luminosity $L_{{\rm max}}$.

Based on a sample of 15 SNe~IIn from PTF/iPTF,
we show that there is a possible correlation
between their rise time and peak luminosity.
Interpreting this correlation in the context of
the relation predicted by Ofek et al. (2014a),
the deduced lower limits on the shock velocity are consistent
with the expected shock velocity from SNe
(i.e., on the order of $10^{4}$\,km\,s$^{-1}$).
Our findings support the suggestion
made by Ofek et al. (2010) and Chevalier \& Irwin (2011)
that the early-time light curves of some SNe~IIn
are powered by shock breakout in a dense CSM.
However, we note that the light curves may be contaminated
by additional sources of energy (e.g., radioactivity),
adding additional spread into the expected relation.
Furthermore, our observations cannot yet be used to rule out
other alternatives (at least not without a detailed model
in hand).

In Figure~\ref{fig:IIn_PeakL_RiseTime} there is a puzzling
deficiency of objects around $L_{0} \approx 10^{45}$\,erg\,s$^{-1}$,
and maybe also some concentration
of events with $L_{0}\approx 4\times10^{44}$\,erg\,s$^{-1}$.
We note that comparison of the $L_{{\rm max}}$ distribution
of our sample and that of 11
SNe~IIn reported by Kiewe et al. (2012)
suggests that this feature may be caused by 
small-number statistics (Fig.~\ref{fig:te_Lp_corr}).
Finally, we propose that application of this test to
other classes of SNe can be used to rule out the hypothesis that they
are powered by interaction of their ejecta with a dense CSM.

\acknowledgments

We thank Dan Perley for obtaining some spectra.
E.O.O. thanks Ehud Nakar and Orly Gnat for discussions.
This paper is based on observations obtained with the
Samuel Oschin Telescope as part of the Palomar Transient Factory
project, a scientific collaboration between the
California Institute of Technology,
Columbia University,
Las Cumbres Observatory,
the Lawrence Berkeley National Laboratory,
the National Energy Research Scientific Computing Center,
the University of Oxford, and the Weizmann Institute of Science.
Some of the data presented herein were obtained at the W. M. Keck
Observatory, which is operated as a scientific partnership among the
California Institute of Technology, the University of California,
and NASA; the Observatory was made possible by the generous
financial support of the W. M. Keck Foundation.  We are grateful for
excellent staff assistance at Palomar, Lick, and Keck Observatories.
E.O.O. is incumbent of
the Arye Dissentshik career development chair and
is grateful to support by
grants from the 
Willner Family Leadership Institute
Ilan Gluzman (Secaucus NJ),
Israeli Ministry of Science,
Israel Science Foundation,
Minerva, Weizmann-UK and
the I-CORE Program of the Planning
and Budgeting Committee and The Israel Science Foundation.
A.G-Y. acknowledge grants from the ISF, BSF, GIF, Minerva,
the EU/FP7 via ERC grant (307260),
and the I-CORE program of the Planning
and Budgeting Committee and The Israel Science Foundation.
M.M.K. acknowledges generous support from the Hubble Fellowship and Carnegie-Princeton Fellowship.
A.V.F.’s SN group at UC Berkeley has received generous financial
assistance from Gary and Cynthia Bengier, the Christopher R. Redlich
Fund, the Richard and Rhoda Goldman Fund, the TABASGO Foundation,
and NSF grant AST-1211916.


\end{document}